# Deep Learning Accelerated Gold Nanocluster Synthesis

Jiali Li[1], Tiankai Chen[1], Kaizhuo Lim[1], Lingtong Chen[2], Saif A. Khan[1], Jianping Xie[1] & Xiaonan Wang[1, *]

[1] Department of Chemical and Biomolecular Engineering, National University of Singapore, 4 Engineering Drive 4, Singapore 117585, Singapore.

[2] Department of Computer Science, University of Southern California, 941 Bloom Walk, Los Angeles, CA 90089, USA.

Jiali Li and Tiankai Chen contributed equally to the work.

[*]Correspondence and requests of materials should be addressed to X. W. (chewxia@nus.edu.sg)

**Abstract**

The understanding of inorganic reactions, especially those far from the equilibrium state, is relatively limited due to their inherent complexity. Poor understandings on the underlying synthetic chemistry have constrained the design of efficient synthesis routes towards desired final products, especially those inorganic materials at atomic precision. In this work, using the synthesis of atomically precise gold nanoclusters as a demonstration platform, we have successfully developed a deep learning framework for guiding material synthesis and accelerating the whole workflow. With only 54 examples, the proposed Graph Convolutional Neural Networks (GCNN) plus Siamese Neural Networks (SNN) classification model with the basic descriptors have been trained. The capability of predicting the target synthesis results has been demonstrated with a successful experimental validation. In addition, understandings in the synthesis process can be acquired from a decision tree trained by a large amount of generated data from the well-trained classification model. This study not only provides a data-driven method accelerating gold nanocluster synthesis, but also sheds light on understanding complex inorganic materials synthesis with low data amount.

**Introduction**

Sub-2 nm gold nanoparticles, or gold nanoclusters (Au NCs), attracted much research interest in the past two decades.[1,2] Undoubtedly, the growing research interest is motivated by the unique properties of Au NCs, such as discrete electronic states, defined molecular formula and structure, quantized charging, molecular chirality, and strong photoluminescence.[3-8] These properties are not observed in bulk gold or gold nanoparticles with core sizes larger than 2 nm. As a result, Au NCs are extensively studied and can be potentially applied in various fields. More interestingly, these properties are greatly determined by the size and structure of the Au NCs, or to be specific, by their molecular formulae. Being ultrasmall in size, Au NCs usually consist of several to one hundred gold atoms in the core, which is stabilized by a shell of ligands (e.g., thiolates and phosphines). Therefore, an atomically precise Au NC can be represented by a formula of $[M_nL_m]^q$, where *n*, *m*, and *q* are the number of metal atoms, ligand molecules, and the net charge in one NC respectively. A small difference in the values of $n$, $m$, and $q$ can greatly affect the properties of the corresponding Au NC.[9] Moreover, the separation of mix-sized Au NCs requires high-resolution separation techniques such as high-performance liquid chromatography due to the inherent tiny difference in their sizes and structures. Hence, obtaining Au NCs at atomic precision after the synthesis will greatly promote the utilization of their properties for potential applications. The synthesis of Au NCs typically adopts from the Brust method.[10] It is a two-step reduction where the ligand is firstly mixed with an Au(III) salt (typically $HAuCl_4$), followed by the addition of another reducing agent (a schematic illustration shown in Figure 1). In the first step, Au(III) is reduced into Au(I), which is then coordinated with the thiolate ligand to form a mixture of Au(I)-ligand complex of various sizes (e.g., $Au_4L_4$, $Au_6L_6$, and $Au_{10}L_{10}$). After the addition of a reducing agent (e.g., $NaBH_4$), a mixture of Au NCs of different sizes will form in the initial stage of reduction. Whether these NC species can grow into a single size (becoming monodispersed at atomic precision) is highly dependent on the size distribution of the initial mixture and the reaction conditions (such process is also known as "size-focusing").[11] As a result, direct synthesis of atomically precise Au NCs is challenging without careful experimental design.

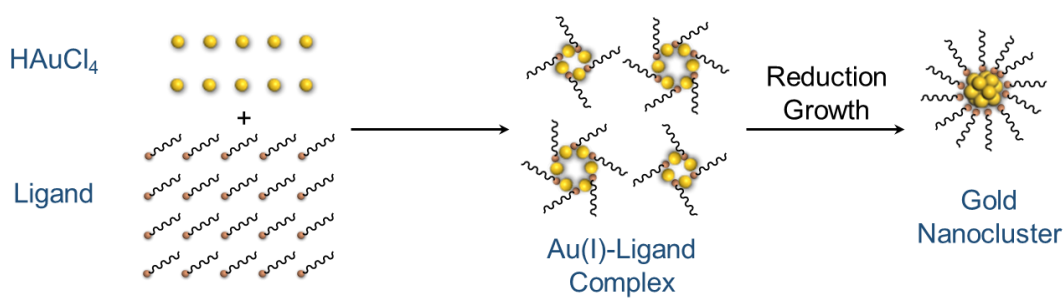

Figure 1 Synthesis of Au NCs. Schematic illustration of the modified Brust method in the synthesis of Au NCs.

Computer-assisted design can accelerate the synthesis process and help chemists develop the chemical synthesis route in a more effective way. From the Materials Genome initiative, first principle simulations along with the quantitative structure-property relationship method as a data-driven tool are built to predict the properties of materials efficiently without having to conduct the actual experiment.[12,13] Based on this effort, more researchers are looking at improving the synthesis process itself. In these studies, both successful and unsuccessful experiments are used to train machine learning (ML) models. These ML approaches do not require solid understanding of a specific domain as required by the first principle simulation approaches. Instead, they are able to gain experience from past experimental data to guide future experiments. Such data-driven screening approaches have been proved successful in organic chemistry, where a large amount of tabulated synthetic data is available.[14-17] Similarly, for inorganic materials, the identification of synthesis parameters driven by ML has also been studied, including the synthesis of a desired phase state and obtaining a certain range of properties.[18-21] However, the synthetic chemistry of the reactions in most of these studies involves only one or limited types of reactions (e.g., with only coordination reactions). On the other hand, the synthesis of Au NCs is more complex. The first stage (mixing of Au(III) salt with ligands) involves with the reduction from Au(III) to Au(I) and the coordination between Au and the ligand, while reduction from Au(I) to Au(0) and the aggregation behavior of Au(0) appear in the second stage reaction. Such a complex reaction system has not yet been explored by ML or other data-driven methods. Nevertheless, ML is able to deal with complex systems in principle from its strong capability of classification tasks in nature. Due to the complexity of the reactions involved in their synthesis, there are very limited number of reports on the direct synthesis of atomically precise Au NCs, resulting in low data amount for the ML training process. Under such conditions, a simple ML model will most likely be unable to learn much information.[22] However, one-shot learning methods such as Siamese Neural Networks (SNN) have potential to perform well with low data amount. This deep neural network has shown strong capability in image recognition with limited examples and promising prediction ability in low-data drug discovery with molecular structure information as an input.[22-24]

Herein, we present a ML model that is able to accelerate the understanding in the synthesis of atomically precise Au NCs under a low data condition by training all the parameters together in this complex reaction system. Although some mechanistic understandings have been gained for the synthetic chemistry, their capabilities to predict the synthesis across the diverse reaction systems are still weak.[25-27] For example, a protocol for the synthesis of atomically precise $Au_{25}$ NCs works only for specific ligands, while the same Au NC species is unlikely to be synthesized when other ligands are adopted. We utilize ML to conduct a classification task based on studying the relationship between reaction conditions, molecular properties, and the final monodispersity in the product (i.e., whether they are atomically precise Au NCs). The ML model uses a SNN stacked with Graph Convolutional Neural Networks (GCNN). With the trained classification model, we can generate new sets of synthetic experimental conditions that are more likely to be successful and can be carried out in the lab, allowing the acceleration of the material discovery process. This first task is also known as synthesis parameters recommendation. In the second part, a "model-of-model" is created to map the black-box SNN onto a human-interpretable decision tree, which can provide more chemical insights despite low amounts of data in the training dataset. With our GCNN + SNN model, we are able to learn insights such as temperature trends, while other models tested in this work failed to learn such insights due to the low data constraint.

## Results

We extracted synthesis conditions of 27 examples from reported literatures and 27 examples from our own lab. The dataset includes synthesis conditions that were able to obtain atomically precise Au NCs in the product (i.e., successful examples, only one Au NC species after the reaction), as well as those not able to obtain atomically precise Au NCs in the product (i.e., unsuccessful examples, more than one Au NC species after the reaction).

To demonstrate the performance of the proposed framework, we have evaluated the synthesis of atomically precise Au NCs in two aspects. The first aspect is the process where the key reaction components (including the ligands, solvent, and reducing agent) for our experiment can be varied. The second aspect is an optimization process of reaction conditions without changing the key reaction components. The former aspect is defined as an explorative process and the latter is defined as an optimization process. To aid in the explorative process, the proposed ML models should be capable to learn the molecular information, which denotes the information based on molecular physicochemical and structure similarity between the key reaction components. Meanwhile, to perform well in the optimization task, the ML models should learn synthetic condition information, which is related to how continuous variations in reaction conditions (e.g., temperature and pH) affect the final results.

As a first attempt, some well-studied machine learning methods, such as Support Vector Machine (SVM) and Dense Neural Network (DNN) were used, with the domain related descriptors as features (Table 1). Such simple strategies haven been proved to be insufficient for this low-data-amount case, as both the key statistical evaluation indicators and probability distributions for experimental conditions are not satisfactory, which will be discussed in detail below. It indicates that a simple ML approach with basic descriptors is not strong enough to learn either molecular information or synthetic condition information effectively for aiding in explorative or optimization process from a small dataset. Thus, a deep learning framework with a closed learning loop based on SNN is constructed as shown in Figure 2. The use of one-shot learning method, SNN, is motivated by the fact that only a limited amount of ground true experiment data in the inorganic material synthesis field is available and the failure in simple ML strategies such as SVM. The aim of this framework is to help material scientists accelerate the understandings in the atomically precise Au NC synthesis through two major approaches: firstly, increasing synthesis and characterization rates of new atomically precise Au NCs by the key classification model; and secondly, mapping the classification model into an interpretable decision-tree to gain chemical insights. The key classification model is a SNN with the GCNN stacked on the top of it. With randomly initialized conditions, the model is able to identify relative promising experimental conditions with high successful probability for carrying out successful experiments with either an explorative or an optimization approach. Outcomes from both successful and unsuccessful experiments can provide feedback to improve this key classification model with a closed learning loop. With the aid of the well-trained key classification model, an arbitrarily large number (about 10,000) of new "synthetic" examples can be generated for building the decision tree, which can provide further information to make the ML model a "white box" with chemical insights. These insights in turn will help in understanding the synthesis process better. The construction of the key classification models, the comparison between different ML methods, and the chemical insights gained from the proposed approach are discussed below.

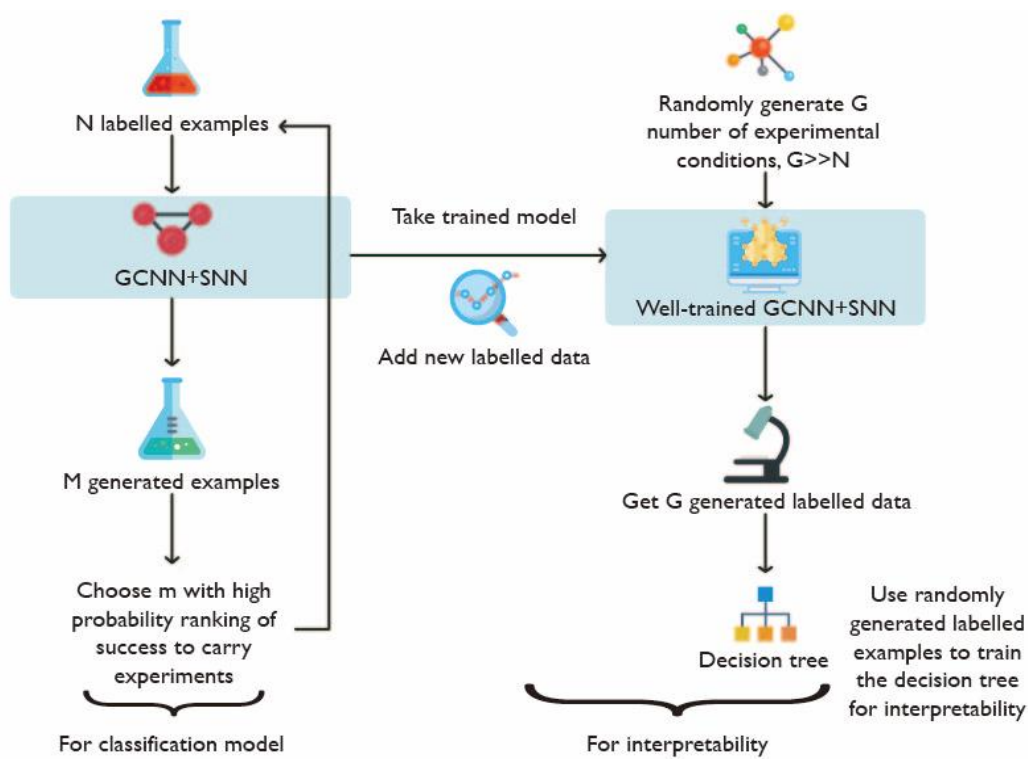

Figure 2 A deep learning framework with a closed learning loop based on SNN. This framework is divided into two parts: 1. Classification model (left), and 2. Interpretation process (right). For the first part: initially, $N$ (a small number in our low data case) labeled examples (i.e., binary classification of atomic precision or not) from literature or lab are collected. These examples are used to train the key classification models, GCNN + SNN in this figure. Although it is feasible to substitute this key classification model with other ML models (e.g., SVM and DNN), the GCNN + SNN model illustrated here is the best performing one which will be discussed later. By using the well-trained model, one can generate $M$ arbitrary number of examples with any arbitrary reaction conditions. These $M$ examples will come out with probability of successful synthesis where scientists can pick the ones with high successful probability to carry out experiments. Both successful and unsuccessful experiments will provide feedback for classification learning to improve the classification performance. For the second part: a large number $G$ (about 10,000) of experimental conditions examples are generated randomly. Then the well-trained classification model in this figure, GCNN + SNN, is taken to predict the outcomes of these examples to label them. These $G$ generated labeled examples are then used to train a decision tree for getting chemical insights.

Key classification model: The dataset we have extracted is heterogeneous from diverse sources. It consists of various examples using different ligands, solvents, and reducing agents aiming to produce Au NCs (such as $Au_{25}$ or $Au_{38}$), providing the molecular information variations. However, it is worthy to note that typically only one or a few reaction conditions are altered from the view of parameter optimization, leading to insufficient information of synthetic condition variations. We labeled an experimental set as 1 if atomically precise Au NCs were obtained in the product and 0 conversely. According to the different features in the 54 examples, two dataset groups were created for training: I, 54 examples from the full dataset (54 dataset I sheet in the SI data); II,

35 examples of aqueous synthesis of $Au_{25}$ NCs (35 dataset II in the SI data). The performance of different ML models on the dataset group I as our key focus is compared. A "model-of-model" is built on the training of the dataset group II to learn the chemical insights based on all possible parameters and especially the effects of both pH and ligand-to-Au molar ratio. It should be noted that the sub-dataset (i.e., dataset group II) is chosen because pH information is not available for synthesis of Au NCs conducted in organic solvents and the ligand-to-Au molar ratio may be in different ranges for the synthesis of differently sized Au NCs.

Table 1 List of basic descriptors and their descriptions. A total of 17 basic descriptors are chosen to represent both the reaction conditions and the key reaction components. 5 reaction condition descriptors are chosen along with 12 descriptors for key reaction components. There are 2 descriptors for describing solvent, 1 descriptor for describing reducing agent and 9 descriptors for describing ligand.[a]

| | Descriptors | | Description of descriptors |
|---|---|---|---|
| Reaction condition descriptors | Ligand concentration | | Concentration of ligand/$HAuCl_4$/reducing agent in the reaction solution |
| | $HAuCl_4$ concentration | | |
| | Reducing agent concentration | | |
| | Reducing reaction temperature | | Temperature of the reaction solution |
| | pH | | pH value of the reaction solution |
| Key reaction components descriptors | Solvent | Dielectric constant | Polarity of the solvent |
| | | Solubility of ligand in corresponding solvent | How well the ligand dissolves |
| | Reducing agent | Redox potential | Ability to reduce |
| | Ligand | Charge potential | Electrostatic polarity |
| | | Hydrogen donor number | Count of hydrogen bond donors |
| | | Hydrogen acceptor number | Count of hydrogen bond acceptors |
| | | Aromatic or aliphatic | With aromatic or aliphatic chain |
| | | Molecular weight | Calculated molecular weight |
| | | Rotatable bonds number | Count of rotatable bonds |

|  |  | Complexity | Measure of structural complexity |
|---|---|---|---|
|  |  | Topological polar surface area | Polar surface area calculated using topological polar surface area method |
|  |  | xlogP3 | Octanol/water partition coefficient calculated using xLogP method[28] |

[a] The references of values of these descriptors are in Descriptors sheet in SI data.

Among the 17 basic descriptors chosen, 12 features are related to the physicochemical properties or the structure information of the key reaction components (e.g., the dielectric constant of the solvent and the number of rotatable bonds of the ligand). These descriptors are used to distinguish the key reaction components (e.g., ligands, solvent, and reducing agent) and find how they affect the final classified state. These descriptors can provide molecular information, which contributes to the ability of the ML models to run explorative process classification. In addition, there are 5 more descriptors on the key operation variables (e.g., temperature and pH) of the whole process, which are related to the optimization process. Apart from these descriptors a GCNN is trained together with classification models in order to account for the rich molecular structure information of the key reaction components. The whole model is illustrated in Figure 3 using SNN with a GCNN stacked on the top.

The SNN is a matching neural network that takes in a pair of input, which is named as an intermediate input vector (IIV) here as shown in Figure 3. Each IIV is passed through an identical half (top or bottom of the SNN) with the same densely connected hidden layer weights and biases. The L1 distance between the last layers of each half of the SNN is taken and connected to one final output node that will be between 0 to 1 in value. A distance value close to 1 means that the two IIV are likely belonging to the same classified state (i.e., both leading to successful synthesis or both leading to unsuccessful synthesis), while the value close to 0 means that they are likely belonging to different classified states. The IIV consists of four parts, three molecular fingerprint vectors (Fp vectors) that contain structure information, and the vector of basic descriptor set (Table 1) which contains other molecular information such as the electronegativity of the ligands. There are one Fp vector for the solvent, the ligand, and the reducing agent, respectively. Those four parts are concatenated together to form the IIV which provides both structural and physicochemical information that a neural network can learn from (rather than discretized molecules as a one-hot vector input to the neural network).

The Fp vector is derived from a GCNN which takes the Simplified Molecular Input Line Entry System (SMILES)[29] canonical name of a molecule as an input. A python library called RDKit converts the SMILES representation into a molecular graph (a set of three tensors) which is input into the GCNN. The input molecular graph (Figure 3 top left in red color with vertices represent individual atoms and edges represent bonds) is convoluted by applying smooth functions (neural networks) to keep track of the information about the substructures between neighbors in each layer (sky-blue lines in Figure 3 molecular graph in red color). To be more specific, this input molecular graph is passed through a hidden convolutional filter (HCF) to form a new layer which has the same graph shape as the input molecular graph but different values at each node. The new convoluted layer is passed through the next hidden convolutional filter to form the next layer and so on. Once the graph passes through all the hidden convolutional filters, each layer will pass through an output convolutional filter (OCF) (left to right direction for top left molecular graph in red color in Figure 3) which transforms the graph to a $D$-dimension vector ($D$ is an arbitrary parameter that can be tuned, in this work $D = 10$) whose length is equal to the Fp vector length. Lastly, the $P + 1$ output vectors ($P$ number of output convolutional filters plus the initial input, in this work $P = 2$) are summed together to give the final Fp vector for that molecule.

After training the GCNN + SNN on a training dataset, the trained model can be used to predict whether an example that has not been observed before belongs to the success or failure class. This is achieved by splitting the training dataset into two support sets, each containing either all the success or failure data respectively. The unseen example is input to the top half of the SNN and an example from the success support set is given to the bottom half of the SNN. This is looped until the unseen example is compared to every other success example from the support set. The output node value is averaged, and a value closer to 1 implies higher probability of success. This is repeated by comparing the unseen example to the failure support set and the output node value is averaged again, and a value closer to 1 implies higher probability of failure this time. If the success support set has a higher average output value than the failure support set, it implies that the unseen example is more likely to be a success case and vice versa.

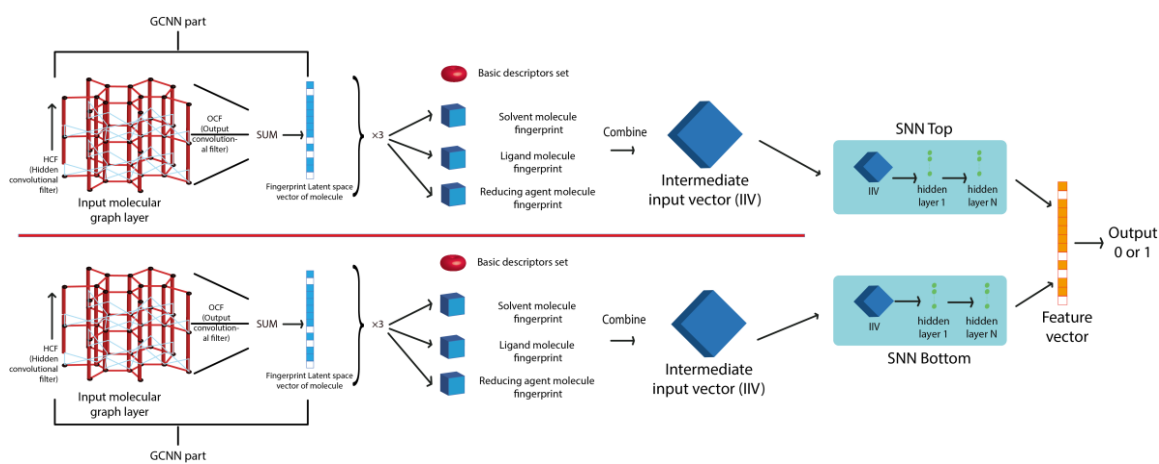

Figure 3 Model illustration. In this paper, GCNN was used to produce Fp for solvent molecules, ligand molecules and reducing agent molecules. The SNN is trained by giving the models batches of pairs from a set of labeled examples (the training dataset) with the half being of the same class and half being a different class. In our model, the GCNN is trained as part of the SNN and its weights and biases are updated together with the SNN instead of being pre-trained. This enables the GCNN to map discrete molecules to their Fp vector latent space in a manner that is more suited for the matching neural network. For SVM and DNN, the input structures are the same as SNN, but have no such separation of top and bottom parts of the classification model.

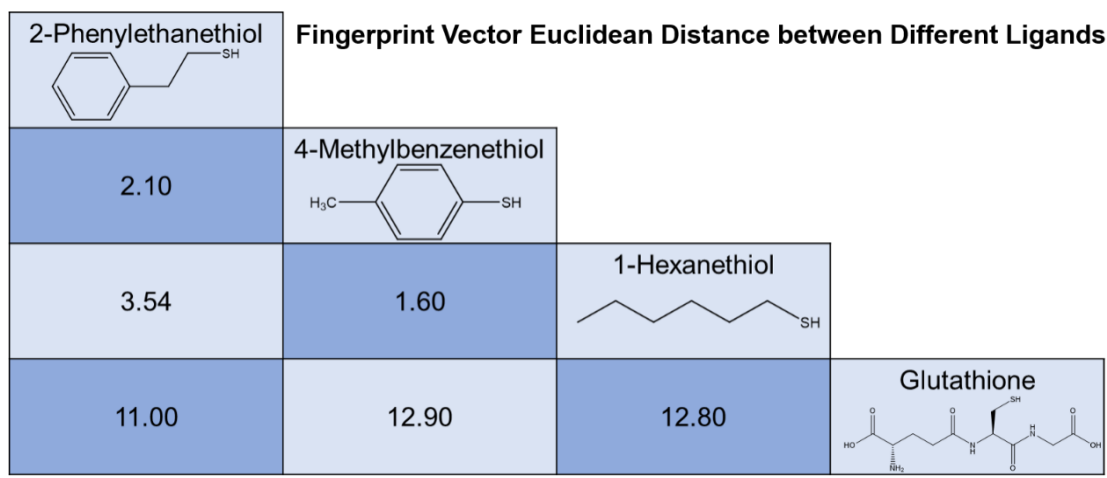

Figure 4 Fingerprint Vector Euclidean Distance calculation. Fingerprint vector Euclidean Distance between four chosen ligands are calculated. If the structures of two ligands are similar, their fingerprint vector output from the GCNN will have a small value of Euclidean Distance between them and vice versa. This shows the capability of GCNN to produce structure similarity information that enhances the performance of classification models by learning molecular information.

To illustrate the ability of the GCNN in providing useful structural information and show it is well trained in our work, we have constructed the adjacency matrix with some sample molecules. Figure 4 shows the Euclidean distance between the calculated vectors of three ligand molecules. The shorter the Euclidean distance between two ligand molecules, the more similar they are in structure and property. This aligns with the chemical similarity between molecules with smaller Euclidean distance. For example, glutathione is a tripeptide featuring functional groups such as amine groups and amide groups, while the other three thiols in Figure 4 are simple thiols with hydrocarbon chains. Hence, glutathione has the longest Euclidean distances with all three other thiols while the Euclidean distances between the three thiols are much shorter. Thus, the GCNN allows the neural network to generalize to unseen set of molecules by considering how similar the unseen molecules are to the set of molecules in the training dataset via the Fp vector output. This is in contrast to training a neural network with molecules that are represented as one-hot vectors (no ability to generalize to new molecules) or with basic descriptors only (might not incorporate as much structural information as a GCNN, see Table 2).

Table 2 Key statistical performance indicators. Performance of six key classification models by using 10-fold validation is described by three statistical indicators. Accuracy is a simple indicator of the proportion of correct predications. The $F_1$ score is the harmonic average of both recall and precision, where an $F_1$ score reaches its best value at 1 and worst at 0. MCC compares the prediction ability of a model to a random guess. If the value is positive it means it is better than random guess. The higher the MCC the better the prediction ability with 1 being the highest value.

| Models | Accuracy | $F_1$ Score | MCC |
| --- | --- | --- | --- |
| GCNN + SNN | 0.81 | 0.79 | 0.65 |
| GCNN + DNN | 0.83 | 0.83 | 0.67 |
| GCNN + SVM | 0.80 | 0.80 | 0.60 |
| SNN | 0.72 | 0.65 | 0.46 |
| DNN | 0.69 | 0.65 | 0.37 |
| SVM | 0.48 | 0.39 | -0.05 |

We have experimented with six different ML classification models, including SVM, DNN, SNN and all of them combined with GCNN, respectively on dataset group I. Table 2 presents the three key statistical performance indicators of the classification models. The Matthews Correlation Coefficient (MCC) is used as the main indicator since it compares the model's predication ability to a random guess. It calculates the Pearson Correlation Coefficient for a two-class confusion matrix and provides a measure of model performance that is unaffected by class imbalance. During the training, 10-fold cross validation is used to validate the model. Since most examples in the dataset feature a different set of key reaction components (lead to molecular information heterogeneity), each fold would likely have a test set that has a different set of reaction components from the training set. Thus, for the classification model to perform well during a 10-fold cross validation, it has to generalize the information it learns from the training set onto the test set with different molecular information, achieving the first goal of performing the explorative task.

From Table 2, the combined GCNN models perform better in all three statistical indicators than models without GCNN. This is likely due to the GCNN ability in providing rich structural information, which enables the model to learn and utilize more molecular information regardless of the type of classification model used. This is important as it potentially allows for an easy method to incorporate molecular structure information by stacking a GCNN at the top and using its output as a feature vector on top of any type of general model. When comparing among the models combined with GCNN, it seems that their performance is comparable to one another. However, the indicators in Table 2 mainly tell the models' ability of molecular information learning and its proficiency in the explorative task. The second goal is to develop a model that can learn synthetic condition information, that is, how the reaction condition affects the probability of a successful synthesis. To investigate whether such learning has been achieved by the six models, we used the trained model as an 'oracle' to generate 10,000 synthetic data, each with the same key reaction component (6-mercaptohexanoic acid protected Au NCs by $NaBH_4$ reduction) but with variations in pH and temperature. The 2D probability map of success for the models are plotted in Figure 5 and in SI Figure S1. The color at each point indicates the probability of success for that particular pH and temperature. Only SNN based models in Figure 5 show variations in probability when the reaction condition varies. This suggests that the SNN based models have learned the synthetic condition information that if the temperature exceeds a certain value (around 50 °C), the probability of successful synthesis of atomically precise Au NCs will diminish to zero as the NCs starts to decompose. This agrees with the conclusion from literature and other experiments.[30] However, for the other four SVM and DNN based models (SI Figure S1 a-b and SI Figure S1 c-d), there is no obvious color variation in the 2D probability map, which indicates no synthetic condition learning achieved by these two models. Lastly, due to the lack of variations in the pH within our dataset, the SNN could not learn as much insights as temperature about how pH affects the synthesis. This makes the probability change not obvious along the pH axis, however there is still pH related probability variation. Apart from analyzing in only two dimensions, a 5D high dimensional visualization is shown in Figure 5c showing the variations in the concentrations of all three key reaction components together with the variation in temperature and pH. This figure shows that our proposed model is capable of investigating the synthesis of atomically precise Au NCs, which is a rather complex system involving a number of variations, through simultaneous multi-dimensional study (a higher dimension can also be achieved but not suitable for visualization here).

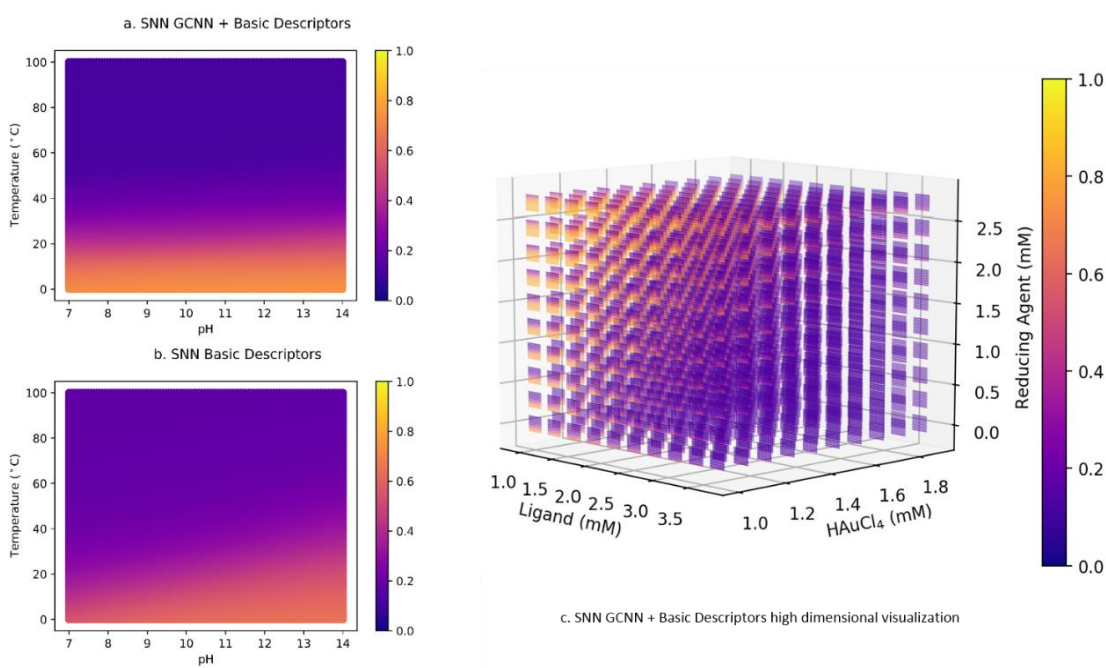

Figure 5 Two 2D probability heatmaps generated from six ML classification models and a 5D probability visualization. For heatmap, the x-axis is pH starting from 7 due to the lack of examples of the synthesis carried in acidic/neutral conditions. The y-axis is temperature with Degree Celsius as the unit. The color-scale indicates the probability of successful synthesis of atomically precise Au NC with yellow as 1 and dark blue as 0. For 5D probability visualization, it shows the variations in the concentrations of all three key reaction components together with the variation in temperature and pH. x, y and z axes are concentrations of ligand, $HAuCl_4$ and reducing agent in mM, respectively. At each point there is a heatmap in pH and temperature similar to the one described above but with less data points (The center of the heat map is the coordinates for concentrations).

In order to validate the reliability of our proposed ML model, we utilized the best performed GCNN + SNN model to predict the probabilities of the synthesis of new synthesis recipes. We validated the performances of the model in both explorative process and optimization process by adopting cysteamine (a ligand did not appear in the dataset) combined with a series of experimental conditions. As shown in the prediction results from SI Table S1, the probabilities of obtaining atomically precise cysteamine-protected Au NCs range from 0.31 to 0.84. We have verified and proved that atomically precise Au NCs can be produced by conducting the synthesis according to the experiment conditions featuring high probabilities. For example, a reaction solution with ligand concentration, HAuCl4 concentration, NaBH4 concentration, and pH at 1 mM, 1.78 mM, 2.33 mM, and 13.22, respectively, is predicted to have a probability of 0.814 to obtain atomically precise cysteamine-protected Au NCs at 22 °C. The UV-Vis absorption spectrum of the as-prepared Au NCs is presented in SI Figure S2. The distinct absorption peaks at 560 nm and 620 nm (as indicated by the arrows in SI) matched well with the reported absorption peaks of $Au_{18}$ NCs, indicating the successful synthesis of $Au_{18}$ NCs at atomic precision following this recipe.[9,26] The success in obtaining an atomically precise Au NC validates our prediction that the probability is high (0.814). To the best of our knowledge, this synthesis protocol to prepare cysteamine-protected $Au_{18}$ has not been reported before, it shows that our GCNN + SNN model is strong in performing both the explorative task and the optimization task.

Deep learning methods such as GCNN + SNN are normally opaque to simple examination as "black-box" models. To gain chemical insights, we developed a "model of the model" by using the best-trained GCNN + SNN as a generative model to generate sufficiently large amounts of synthetic data followed by using the synthetic data to train a decision tree. Firstly, the GCNN + SNN is trained and used to generate a large number of random examples (11,095 in total, as a random number around 10,000) by initializing all the descriptors randomly. We generated these examples because the original dataset is too small to train a decision tree, but this can be overcome by generating a large synthetic dataset. Although this decision tree model will perform no better than the SNN model itself, the interpretations and implications can play an important role in understanding the Au NCs synthesis and accelerating the domain development. As a proof of concept, we utilized the decision tree approach to generate chemical insights in synthesis of atomically precise $Au_{25}$ NCs in aqueous phase. Thus, dataset group II is adopted here. The calculated $F_1$ score and MCC value of the decision tree are 0.95 and 0.90, respectively, calculated by evaluating another 1,000 randomly generated examples from the GCNN + SNN. Such high $F_1$ test value and MCC value indicate that the building of a "model of model" is promising as the decision tree well maps the well-trained "black-box" SNN model.

The decision tree is shown in the flowchart in Figure 6. From this flowchart we can generate some synthetic chemistry guidelines to assist in designing the synthesis route. The decision tree examines the five reaction conditions in the synthesis of Au NCs in aqueous phase including the chain of the ligand (aromatic or aliphatic), the ligand to Au molar ratio, the concentration of reducing agent, pH of the reaction solution, and the reaction temperature. The probabilities of the successful synthesis are given based on the combination of all five conditions. The tree shows that firstly, when water is used as the solvent for the synthesis of Au NCs, using aliphatic ligands will have much higher chance of obtaining atomically precise Au NCs compared with using aromatic ligands. This is consistent with the fact that aromatic ligands are overall less soluble in water, and good ligand solubility is critical for a well-controlled synthesis. Secondly, the ligand-to-Au molar ratio is found to be an important factor for the aqueous phase synthesis of $Au_{25}$ NCs. The model predicts that for successful synthesis, the ligand-to-Au molar ratio should be less than 6.0. The knowledge again matches with our understandings that Au(I)-ligand complexes of different size and structure will form at different ligand-to-Au molar ratios. The short Au(I)-ligand complexes formed at high ligand-to-Au molar ratios favors the formation of large Au nanoparticles instead of NCs.[12] Moreover, the effects of the reducing agent concentration have been learned and investigated. On one hand, if the reducing agent concentration falls below 52 mM, the reaction temperature should be kept below 50 °C for the formation of atomically precise $Au_{25}$. This is probably because at low concentrations of reducing agent, low reaction temperature preventing the decomposition of Au NCs is more important than in the mild reduction environment.[30] On the other hand, a pH value below 12.8 is found critical for the atomically precise $Au_{25}$ NC synthesis in the cases of reducing agent concentration above 52 mM. It should be noted that only alkaline conditions (pH > 7.0) have been trained for the pH values due to the lack of examples of the synthesis carried in acidic/neutral conditions. The weakly alkaline condition is important for simultaneously tuning the formation kinetics and thiol etching abilities in the growth of $Au_{25}$ NCs.[31]

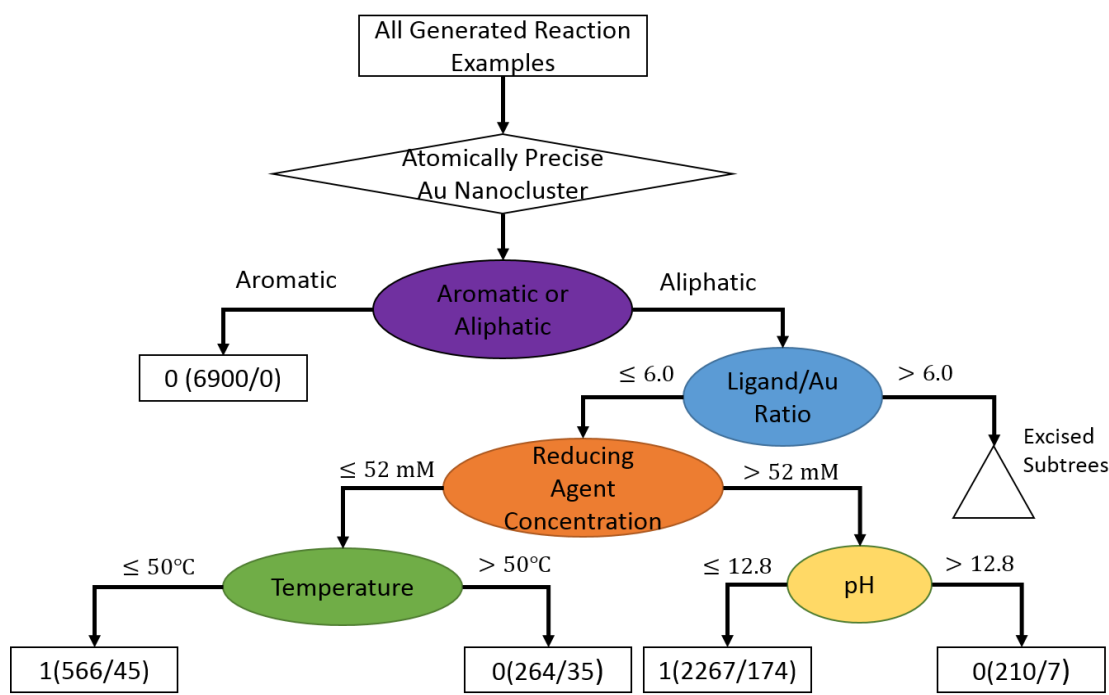

Figure 6 Decision tree. Ovals represent decision nodes and rectangles represent reaction-outcome bins. Triangles mean excised subtrees due to both extra small examples in that branch and chemical intuition. The numbers on the arrows correspond to decision attributing test values. Each reaction-outcome bin (rectangle) corresponds to a specific reaction-outcome value (1 success, 0 failure). The number in parentheses is the number of reaction examples correctly assigned to that bin (any incorrectly classified reactions are given after a slash).

**Discussion**

To conclude, we have shown machine learning accelerates the synthesis of atomically precise Au NCs by incorporating all parameters in the synthesis into consideration instead of focusing on one or two parameters in the experimental discoveries. Our purely data-driven results show that there is a quantitative probability pattern of successful synthesis of atomically precise Au NCs based on the combinations of reagents and the reaction conditions. Moreover, the probability pattern predicts the synthesis of Au NCs protected by cysteamine, a ligand absent from the dataset, and the prediction is experimentally validated by the successful synthesis of $Au_{18}$ NCs. Within this two-step machine learning framework, chemical insights for this complex reaction system (comprising of various types of reactions towards an atomically precise product) have been effectively generated. The combinations of a few key parameters including ligand type, ligand-to-Au molar ratio, pH, and reaction temperature have been identified for the successful synthesis of atomically precise Au NCs from the model-of-model decision tree. In addition, our GCNN + SNN approach works well with low data (only 54 examples, instead of >1000 typically used in ML studies on chemical synthesis), which is a common situation where obtaining the desired product is challenging. Our work has provided a framework of classifying the diverse parameters in a chemical synthesis and elucidated machine learning applications in a complex chemical synthesis system with limited number of successful examples.

Despite the good explorative process performance of the proposed model, the performance in optimizing the reaction conditions needs further development. Currently, from the heterogeneous property of the available dataset, the key reaction components that leading to successful synthesis of atomically precise Au NCs can be identified. However, predicting the exact composition in the atomically precise Au NCs is not maturely available yet. This is due to lack of both successful and unsuccessful examples for machines to classify. We anticipate the prediction ability and interpretation ability of the proposed framework with available models can be improved by further high-throughput experiments focused on a specific system with fixed Au atom number. We identify this as a promising direction for future investigation.

**Methods**

**Machine learning.** The training of the six key classification models is constructed by using Keras library. The GCNN is implemented by using Keras Neural Graph Fingerprint codes and Rdkit (https://github.com/rdkit/rdkit). SVM and DNN are well-known models so here we briefly introduce our SNN model.

The Graph Convolutional Neural Network takes the reaction molecules as inputs and gives a fingerprint vector V. It is then concatenated with the vector W representing the selected features from the reaction using a direct sum of the vector space to produce the intermediate input vector (IIV) X.

$$X = V \oplus W \qquad (1)$$

Pairs of IIV is then put into the Siamese Neural Network. This model learns a measure

of similarity. Assume we have input X, for a half model parameterized by *W*, it returns result $G_w(X)$. The similarity metric for a pair of input $X_1$ and $X_2$ is then: [27]

$$E_w(X_1, X_2) = ||G_w(X_1) - G_w(X_2)|| \qquad (2)$$

which is the L1 distance between the two outputs. The parameters *W* of the model needs to be trained so that if $X_1$ and $X_2$ are in the same class (both success or failure), the similarity metric is small and vice versa.

Therefore the loss function must be have contrastive terms for the input pair with the same classes and the different classes. We used binary cross entropy function and the general term for *N* inputs can be written as:

$$\mathcal{L} = \frac{1}{N} \sum_{i=1}^{N} L(W, (Y, X_1, X_2)^i) \qquad (3)$$

$$L(W, (Y, X_1, X_2)^i) = (1-Y)log(E_w(X_1, X_2)) + Ylog(1 - E_w(X_1, X_2)) \qquad (4)$$

Here $(Y, X_1, X_2)^i$ is the *i*-th example with inputs $X_1$, $X_2$ and label *Y*. *Y* is 1 if the inputs are in the same class or 0 if they are in different classes. The log function is a smooth monotonically increasing function so that the minimization of the loss function would maximize the L1 distance between the pair in different classes and minimize it for the same class pairs.

The hyperparameters of these models are optimized by running 50 trials of Gaussian process optimization (gp minimize) using scikit opt library. The goal is to minimize MCC. The best hyperparameter based on the lowest MCC value out of the 50 trial runs is chosen (details in SI). The fingerprint Euclidean distance is calculated by taking GCNN portion of the trained model. The simplified molecular-input line-entry system code of selected molecules are transferred by Rdkit into graph tensor and then input into the GCNN portion to get the fingerprint vector output. Since we used 10-fold cross validation, one trained model has 10 instances each with a different training dataset. Repeat the fingerprint calculation steps for all 10 instances and take the average output. The norms between the various ligands to get the Euclidean distance adjacency matrix is taken. The 2D probability heatmap is constructed by repeating the following steps for all six key classification models. 10,000 new examples using a mesh grid for pH and temperature are generated while the other experimental conditions and key reaction components are kept constant. Input the 10,000 examples into the trained model to get an evaluated output. Repeat the above two steps for all 10 instances from 10-fold validation and take the average output. Plot a scatter plot for the 10,000 points with the pH and temperature as x and y axis and the colorscale as the probability of success. For 5D visualization the same approach is adopted, however in each dimension, 10 evenly spacing coordinates in a reasonable range similar to the experimental conditions are chosen (e.g. for pH we chose 7 to 14) and these synthesized experimental conditions are predicted by the trained model. The three space coordinates represents the concentration of a reaction component and at each point a heat map similar to the one mentioned above is plotted. The details of experimental validation prediction method is in SI. The model-of-model decision tree is constructed by randomly generating about 10,000 examples varying all variables. Input the generated examples into a trained model to classify into success or failure and use the output examples to train the decision tree using sklearn library with details in SI.

**Synthesis of Gold Nanoclusters**. The synthesis of gold nanoclusters is modified from Brust method. In general, a ligand solution was first added to the solvent, followed by the addition of $HAuCl_4$ solution. After that, the pH of the solution was adjusted to the desired value (for aqueous phase synthesis). A reducing agent solution is then mixed with the reaction mixture for the reduction into gold nanoclusters. The detailed parameters for the synthesis of lab examples are listed in SI data.

**Supplementary Information** is available in the online version of the paper.

**Codes availability**

All codes are available on GitHub via
https://github.com/amdprojectwanggroup/AMD-Project.git

**Reference**

1. Chakraborty, I. & Pradeep, T. Atomically Precise Clusters of Noble Metals: Emerging Link between Atoms and Nanoparticles. *Chem. Rev.* **117**, 8208-8271, (2017).
2. Jin, R. C., Zeng, C. J., Zhou, M. & Chen, Y. X. Atomically Precise Colloidal Metal Nanoclusters and Nanoparticles: Fundamentals and Opportunities. *Chem. Rev.* **116**, 10346-10413, (2016).
3. Venzo, A. *et al.* Effect of the Charge State (z =-1, 0, +1) on the Nuclear Magnetic Resonance of Monodisperse $Au_{25}[S(CH_2)_2Ph]^z$ Clusters. *Anal. Chem.* **83**, 6355-6362, (2011).
4. Jadzinsky, P. D., Calero, G., Ackerson, C. J., Bushnell, D. A. & Kornberg, R. D. Structure of a Thiol Monolayer-protected Gold Nanoparticle at 1.1 Angstrom Resolution. *Science* **318**, 430-433, (2007).
5. Dolamic, I., Knoppe, S., Dass, A. & Burgi, T. First enantioseparation and circular dichroism spectra of $Au_{38}$ clusters protected by achiral ligands. *Nat. Commun.* **3**, 798, (2012).
6. Luo, Z. T. *et al.* From Aggregation-Induced Emission of Au(I)-Thiolate Complexes to Ultrabright Au(0)@Au(I)-Thiolate Core-Shell Nanoclusters. *J. Am. Chem. Soc.* **134**, 16662-16670, (2012).
7. Yu, Y. *et al.* Identification of a Highly Luminescent $Au_{22}(SG)_{18}$ Nanocluster. *J. Am. Chem. Soc.* **136**, 1246-1249, (2014).
8. Aikens, C. M. Electronic Structure of Ligand-Passivated Gold and Silver Nanoclusters. *J. Phys. Chem. Lett.* **2**, 99-104, (2011).
9. Negishi, Y., Nobusada, K. & Tsukuda, T. Glutathione-protected Gold Clusters Revisited: Bridging the Gap between Gold(I)-thiolate Complexes and Thiolate-protected Gold Nanocrystals. *J. Am. Chem. Soc.* **127**, 5261-5270, (2005).
10. Brust, M., Walker, M., Bethell, D., Schiffrin, D. J. & Whyman, R. Synthesis of Thiol-derivatized Gold Nanoparticles in a 2-phase Liquid-liquid System. *J. Chem. Soc., Chem. Commun.*, 801-802, (1994).
11. Jin, R. C. *et al.* Size Focusing: A Methodology for Synthesizing Atomically Precise Gold Nanoclusters. *J. Phys. Chem. Lett.* **1**, 2903-2910, (2010).
12. Le, T., Epa, V. C., Burden, F. R. & Winkler, D. A. Quantitative structure-property relationship modeling of diverse materials properties. *Chem. Rev.* **112**, 2889-2919, (2012).
13. Kalil, T. & Wadia, C. Materials Genome Initiative for Global Competitiveness. *Technical Report, National Science and Technology Council*, (2011).
14. Grzybowski, B. A., Bishop, K. J. M., Kowalczyk, B. & Wilmer, C. E. The 'wired' universe of organic chemistry. *Nat. Chem.* **1**, 31-36, (2009).
15. Szymkuć, S. *et al.* Computer-Assisted Synthetic Planning: The End of the Beginning. *Angew. Chem. Int. Ed.* **55**, 5904-5937, (2016).

16  Ley, S. V., Fitzpatrick, D. E., Ingham, R. J. & Myers, R. M. Organic synthesis: March of the machines. *Angew. Chem. Int. Ed.* **54**, 3449-3464, (2015).
17  Coley, C. W., Barzilay, R., Jaakkola, T. S., Green, W. H. & Jensen, K. F. Prediction of Organic Reaction Outcomes Using Machine Learning. *ACS Cent. Sci.* **3**, 434-443, (2017).
18  Raccuglia, P. *et al.* Machine-learning-assisted materials discovery using failed experiments. *Nature* **533**, 73-76, (2016).
19  Copp, S. M. *et al.* Fluorescence Color by Data-Driven Design of Genomic Silver Clusters. *ACS Nano* **12**, 8240-8247, (2018).
20  Balachandran, P. V., Kowalski, B., Sehirlioglu, A. & Lookman, T. Experimental search for high-temperature ferroelectric perovskites guided by two-step machine learning. *Nat. Commun.* **9**, 1668, (2018).
21  Chopra, S., Hadsell, R. & LeCun, Y. in *2005 IEEE Computer Society Conference on Computer Vision and Pattern Recognition (CVPR'05).*  539-546 vol. 531.
22  Altae-Tran, H., Ramsundar, B., Pappu, A. S. & Pande, V. Low Data Drug Discovery with One-Shot Learning. *ACS Cent. Sci.* **3**, 283-293, (2017).
23  Koch, G., Zemel, R. & Salakhutdinov, R. Siamese neural networks for one-shot image recognition.  in *ICML Deep Learning Workshop.*   (2015).
24  Duvenaud, D. *et al.* Convolutional Networks on Graphs for Learning Molecular Fingerprints. in *Advances in Neural Information Processing Systems.* **28** 2224-2232 (2015).
25  Chen, T. K., Luo, Z. T., Yao, Q. F., Yeo, A. X. H. & Xie, J. P. Synthesis of Thiolate-protected Au Nanoparticles Revisited: U-shape Trend between the Size of Nanoparticles and Thiol-to-Au Ratio. *Chem. Commun.* **52**, 9522-9525, (2016).
26  Yu, Y. *et al.* Scalable and Precise Synthesis of Thiolated $Au_{10-12}$, $Au_{15}$, $Au_{18}$, and $Au_{25}$ Nanoclusters via pH Controlled CO Reduction. *Chem. Mater.* **25**, 946-952, (2013).
27  Parker, J. F., Weaver, J. E. F., McCallum, F., Fields-Zinna, C. A. & Murray, R. W. Synthesis of Monodisperse $Oct_4N^+$ $Au_{25}(SR)_{18}^-$ Nanoparticles, with Some Mechanistic Observations. *Langmuir* **26**, 13650-13654, (2010).
28  Cheng, T. *et al.* Computation of Octanol - Water Partition Coefficients by Guiding an Additive Model with Knowledge. *Journal of Chemical Information and Modeling* **47**, 2140-2148, (2007).
29  Weininger, D. SMILES, a chemical language and information system. 1. Introduction to methodology and encoding rules. *Journal of Chemical Information and Computer Sciences* **28**, 31-36, (1988).
30  Chen, T. K., Yao, Q. F., Yuan, X., Nasaruddin, R. R. & Xie, J. P. Heating or Cooling: Temperature Effects on the Synthesis of Atomically Precise Gold Nanoclusters. *J. Phys. Chem. C* **121**, 10743-10751, (2017).
31  Yuan, X. *et al.* Balancing the Rate of Cluster Growth and Etching for Gram- Scale Synthesis of Thiolate- Protected $Au_{25}$ Nanoclusters with Atomic Precision. *Angew. Chem. Int. Ed.* **53**, 4623-4627, (2014).

**Acknowledgments**

We acknowledge the Singapore RIE2020 Advanced Manufacturing and Engineering (AME) Programmatic grant "Accelerated Materials Development for Manufacturing". The authors thank Y. Zhang in designing Figure 2 and 3.

**Author contributions**
J.L. and T.C. developed the database. J.L, K.L. and L.C. developed the machine learning models and statistical analyses, supervised by X.W.. T.C. performed the experimental preparation and validation, supervised by J.X.. S.A.K., J.X. and X.W conceived and executed the project. J.L. and T.C. wrote the manuscript with input from all the co-authors. All authors discussed the results and commented on the manuscript.

Competing financial interests: The authors declare no competing financial interests.